\documentclass[fleqn,10pt]{wlscirep}
\usepackage{graphicx}
\usepackage{subfigure}
\usepackage{float}
\title{Magnetic Skyrmions for Cache Memory}
\author[1]{Mei-Chin Chen}
\author[1]{Kaushik Roy}
\affil[1]{School of Electrical and Computer Engineering, Purdue University, West Lafayette, 47906, USA}
\affil[*]{chen1320@purdue.edu}
\begin{abstract}
Magnetic skyrmions (MS) are particle-like spin structures with whirling configuration, which are promising candidates for spin-based memory. MS contains alluring features including remarkably high stability, ultra low driving current density, and compact size. Due to their higher stability and lower drive current requirement for movement, skyrmions have great potential in energy efficient spintronic device applications. 
We propose a skyrmion-based cache memory where data can be stored in a long nanotrack as multiple bits. Write operation (formation of skyrmion) can be achieved by injecting spin polarized current in a magnetic nanotrack and subsequently shifting the MS in either direction along the nanotrack using charge current through a spin-Hall metal (SHM), underneath the magnetic layer. The presence of skyrmion can alter the resistance of a magnetic tunneling junction (MTJ) at the read port. Considering the read and write latency along the long nanotrack cache memory, a strategy of multiple read and write operations is discussed. Besides, the size of a skyrmion affects the packing density, current induced motion velocity, and readability (change of resistance while sensing the existence of a skyrmion). Design optimization to mitigate the above effects is also investigated.    
\end{abstract}
\begin{document}
\flushbottom
\maketitle
\thispagestyle{empty}
\section*{Introduction}
Leakage current and process parameter variations significantly impact scaled CMOS (complementary metal-oxide semiconductor) devices. The need for non-volatility (zero off-state leakage), higher density, and robustness has consequently led researchers to explore alternatives technology to replace traditional CMOS-based memories. Several emerging technologies such as phase change memory (PCM), resistive random-access memory (RRAM), spin-transfer torque Magnetic RAM (STT-MRAM), and domain wall motion (DWM) based memory have been proposed as potential substitutes for CMOS-based memories. 
One such promising high density memory technology, domain wall motion based racetrack memory, was proposed by IBM\cite{IBM_racetrack}. In such a racetrack memory, multiple data bits can be coded in a sequence of spin ups or downs, separated by domain wall, and is driven by the spin transfer torque. DWM based TapeCache \cite{tapecache} has shown good performance improvement (with higher packing density and better energy efficiency) over other spintronic memory devices \cite{nucleation}. However, the motion of domain wall might be pinned by the presence of defects\cite{defect_in_DWM}, and hence the feasibility of the DWM based memory might be limited by the imperfectness in the material. One attractive alternative is magnetic skyrmion. 
Skyrmions, as magnetic storage, have been shown to possess few benefits over the domain wall motion based racetrack memory in terms of high density, low power, and is less limited by imperfectness of the material. Topologically protected properties prevents the motion of skyrmions from pinning at the defect sites in a magnetic layer, and thus skyrmions as information carrier are robust and more resistant to pinning defects. Skyrmions can be observed in non-centrosymmetric bulk magnetic materials or ultrathin magnetic system with breaking inversion symmetry and large spin orbital coupling. The state of a magnetic skyrmion can be explained by the presence of Dzyaloshinskii-Moriya Interaction (DMI)\cite{dzyaloshinskyDMI,moriyaDMI} -- the DMI between two atomic spins $S_1$ and $S_2$ with an neighboring atom\cite{fert2013skyrmions,nagaosa2013topological,muhlbauer2009skyrmion,yu2010real,yu2011near,heinze2011spontaneous} can be expressed as 
\begin{equation}
H_{DM}=-D_{1,2}\cdot (S_1\times S_2)
\end{equation}
where $D_{1,2}$ is the Dzyaloshinskii-Moriya (DM) vector. \par
\begin{figure}[ht]
\centering
\includegraphics[width=0.8\linewidth]{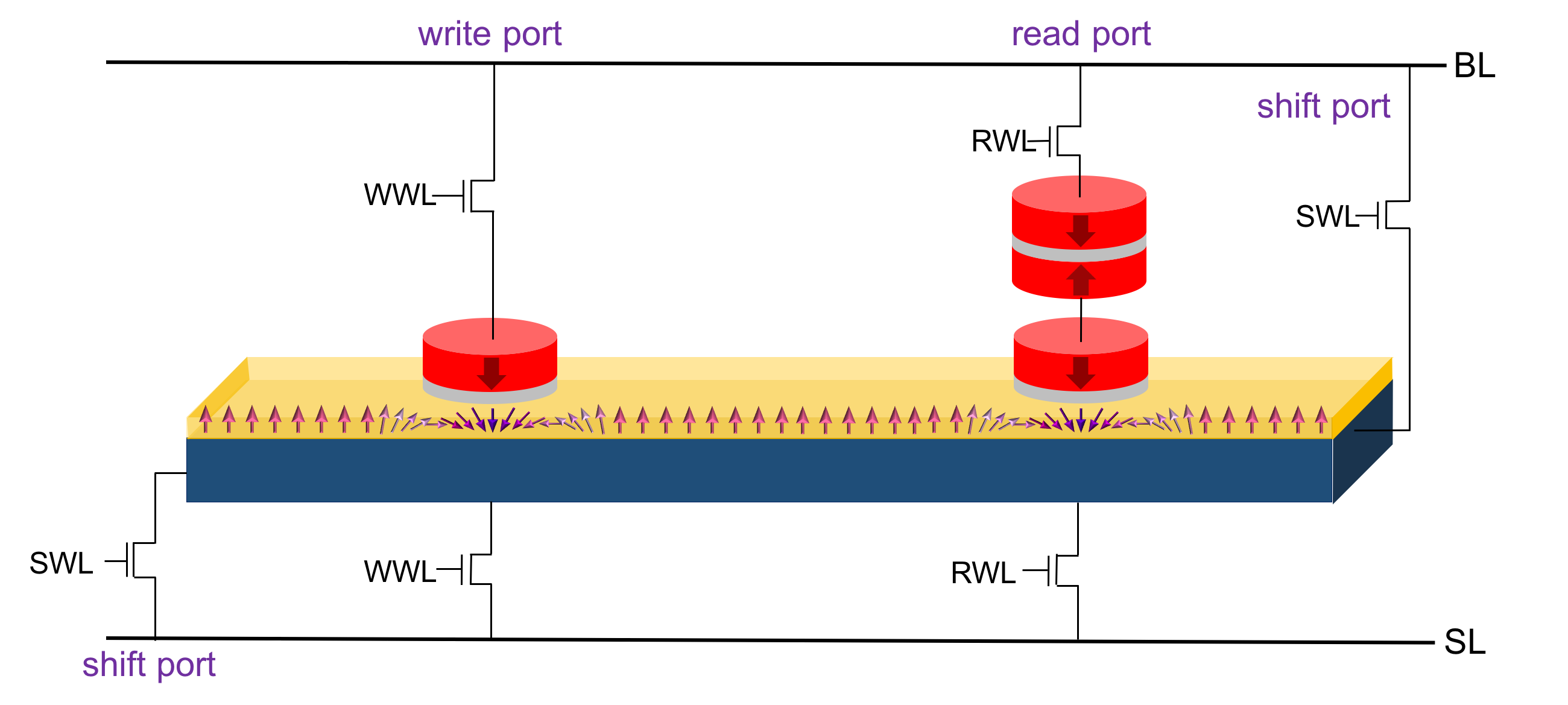}
\caption{\textbf{Schematic of device.} The write/shift/read operations can be performed by the proposed device structure. A skyrmion can be nucleated in the nanotrack (yellow layer) by injecting a spin-polarized current through the left MTJ. The motion of skyrmions can be driven by utilizing vertical injection of spin current generated from a charge current flowing through the spin-Hall Metal layer (blue layer). The presence of a skyrmion can be detected by sensing the change of tunneling conductance of the right MTJ with respect to a reference MTJ. Here, the existence of a skyrmion represents logic “1”, while its absence denotes logic “0”}
\label{fig:Device_Structure}
\end{figure}

 
In this work, we propose a skyrmion based on-chip cache memory. The much higher density of the skyrmion based cache memory and its non-volatility are promising for last level on-chip cache application. 
Fig. \ref{fig:Device_Structure} shows our proposed device structure of a multi-bit skyrmion cell. Magnetic skyrmions can be stored in a ferromagnetic nanotrack adjacent to an SHM. 
The write port consists of an MTJ with two access transistors. A skyrmion can be nucleated in the nanotrack by injecting a spin-polarized current through the left MTJ (write MTJ)\cite{nucleation}. 
The motion of skyrmions can be controlled by an in-plane current through the nanotrack, or by utilizing vertical injection of spin current generated from a charge current flowing through the spin-Hall Metal layer (SHM)\cite{liu2012spin,finocchio2013switching}. It has been shown that skyrmions driven by current flow through the SHM layer can obtain higher velocities with lower current densities\cite{nucleation}. Consequently, we use a charge current through the SHM layer to move the skyrmions along the current flow direction. The shift ports at both sides of the SHM consists of an access transistor. Stored data can be moved in either direction by turning ON the access transistor and by properly biasing BL and SL. The presence of a skyrmion can be detected by sensing the change of tunneling conductance of the right MTJ (read MTJ) with respect to a reference MTJ and a simple CMOS inverter. The read port includes an MTJ in series with a reference MTJ and two access transistors. The presence of a skyrmion under the read port alter the resistance of the right MTJ. This resistance change can be sensed by the voltage difference between the read MTJ and the reference MTJ. 
Note, however, the motion of skyrmions might bend away from the intended direction due to the Magnus force \cite{thiele1973steady}. Also, the repulsive force from neighboring skyrmions might cause the distance between the data bits to be inconsistent. To correctly sense the stored bits, it is necessary to move the target skyrmion bit to a position right underneath the read head. We investigate robust operation of memories by ensuring proper motion of skyrmions along the center of the nanotrack, and alleviate the influence of repulsive force from neighboring skyrmions by reliable spacing between consecutive bits. 
We also investigate design optimization. The size of skyrmions influences the packing density of the proposed cache memory, current induced motion along the nanotrack, and the resistance change while detecting the presence of a skyrmion. The skyrmion size can be tuned by intrinsic parameters (such as Dzyaloshinskii–Moriya interaction strength and perpendicular magnetic anisotropy constant) and also by the width of the nanotrack.

\begin{figure}[ht]
\centering
\includegraphics[width=0.8\linewidth]{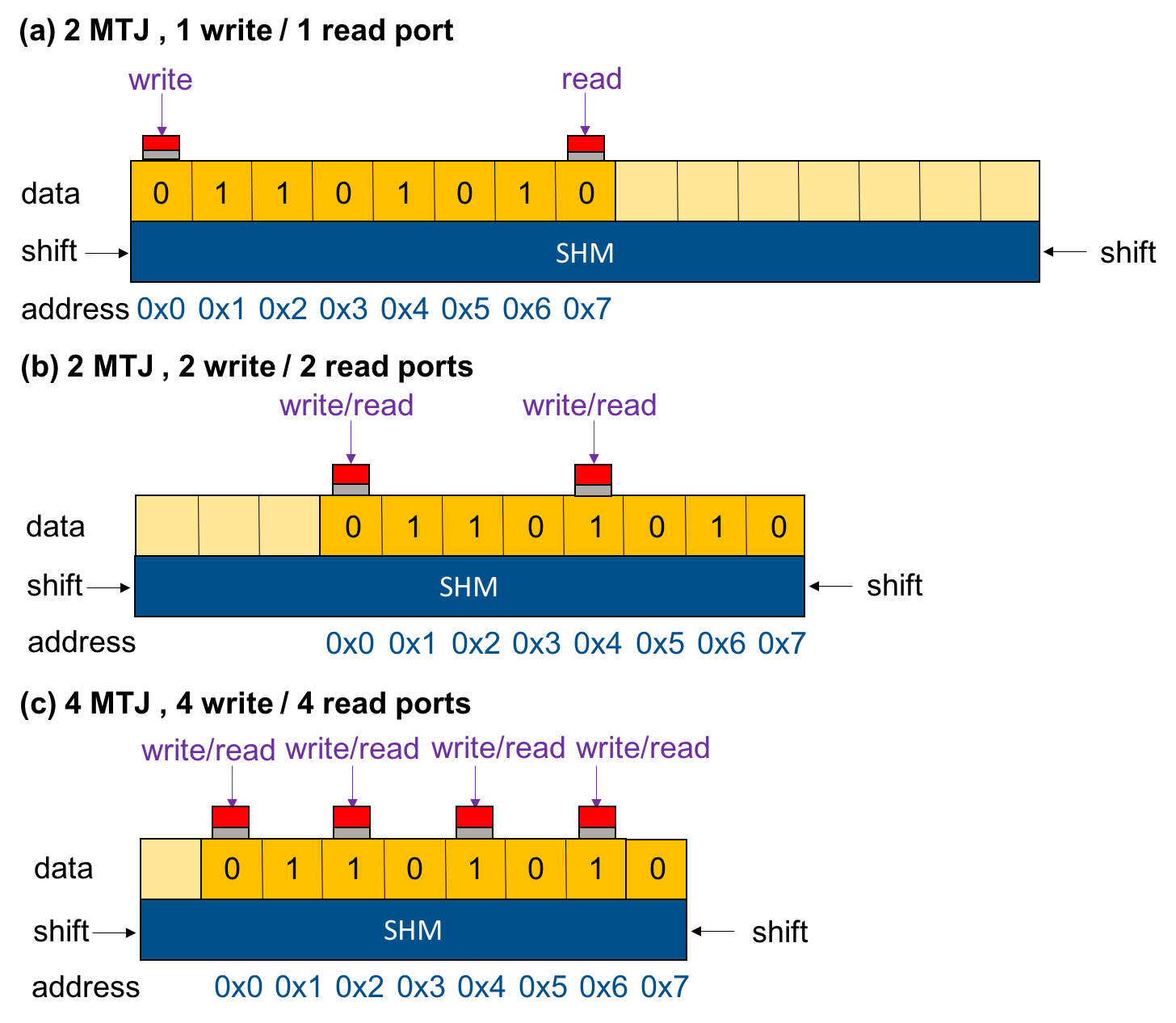}
\caption{Logic view of multi-bit skyrmion based cache memory}
\label{fig:schematic_operation}
\end{figure}
\begin{figure}[ht]
\centering
\includegraphics[width=0.8\linewidth]{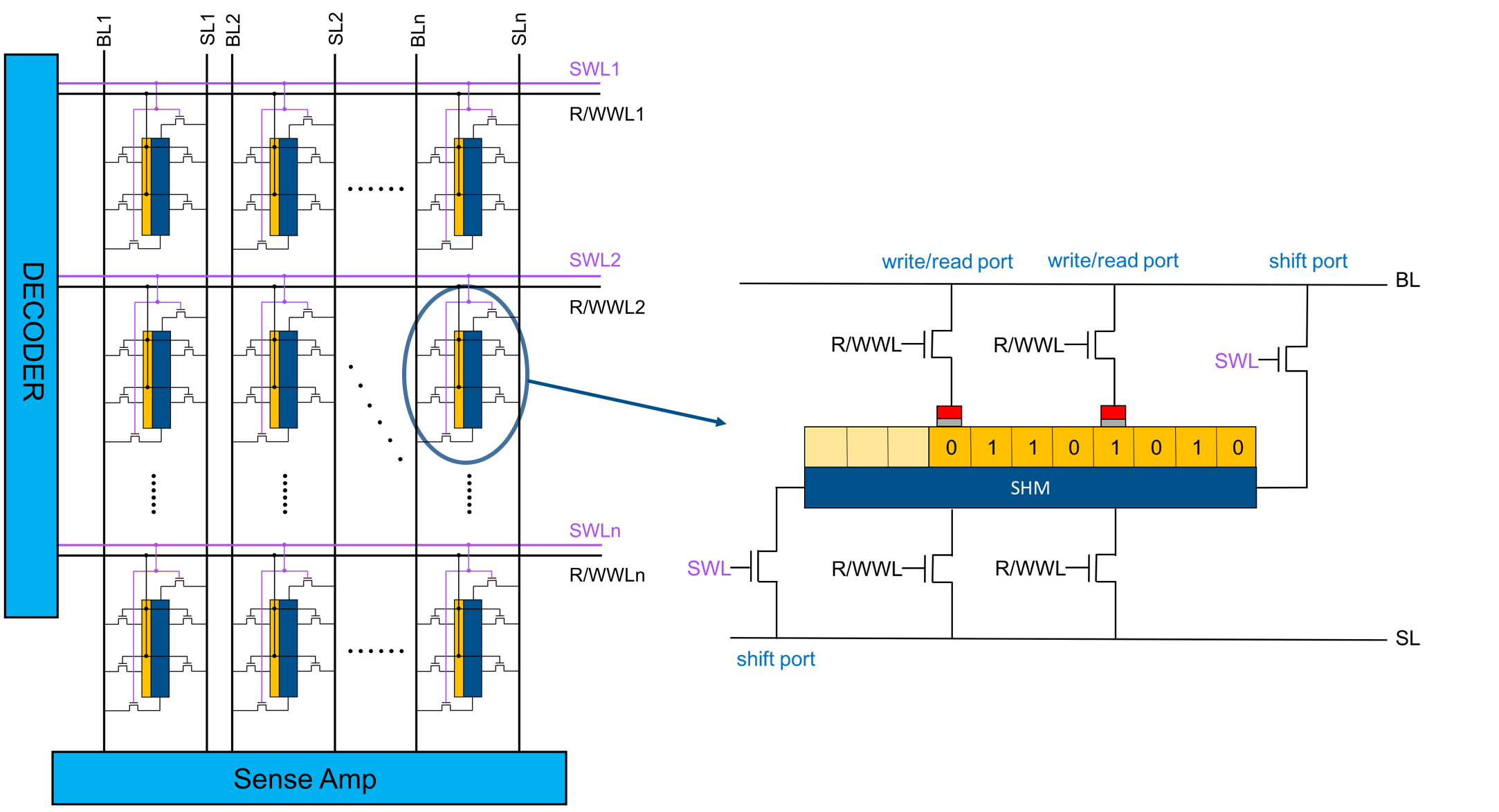}
\caption{Organization of skyrmion based multi-bit cells in a memory array.}
\label{fig:system}
\end{figure}
\section*{Results}
\subsection*{Operation of multi-bit skyrmion cell}
In an ultrathin ferromagnetic nanotrack with strong spin-orbital coupling (SOC) and broken inversion symmetry, large DMI at the interface of nanotrack and the SHM stabilizes the presence of skyrmion. 
Fig. \ref{fig:schematic_operation} shows the logical representation of stored data along the nanotrack. Depending on the existence of a skyrmion, different logic values can be stored along the nanotrack as multiple bits. Existence of a skyrmion represents logic “1”, while its absence denotes logic “0” at the corresponding address. 
A current injected into the SHM (blue layer) from the right can shift skyrmions in the nanotrack right and vice versa. The device structure of a word with single port or multiple write/read ports are shown in Fig. \ref{fig:schematic_operation}(a) and Fig. \ref{fig:schematic_operation}(b), \ref{fig:schematic_operation}(c), respectively. 
Note that the write and read MTJs can be placed at any address location. As an example, consider Fig. \ref{fig:schematic_operation}(a) which shows a write MTJ at address "0x0", a read MTJ at address "0x7" and we write a sequence of 0's and 1's. During the first write cycle, "0" is written into the address "0x0", and subsequently shifted right to the next address "0x1". "1" is written into the address "0x0" during the next write cycle, and then the stored data in the nanotrack is shifted right to the next address. By repeatedly writing data into the address "0x0" and shifting all stored data right to the next address, a sequence of bits can be stored in the nanotrack. 
As shown in Fig. \ref{fig:schematic_operation}(a), the read MTJ is located on address "0x7". To read the stored data at address "0x4" ,say, the bit is shifted right by three positions to the address under the read MTJ. However, in order to prevent stored data in the nanotrack being destructed during shift operation, we extend the nanotrack by having extra data bits. 
In the worst case, to read the stored data at address "0x0", the bit is required to be shifted right by seven positions before it can be sensed. Note that the write/read latency is equal to the length of the word.
However, we can also modify the write/read latency by introducing multiple write/read ports, allowing simultaneous read and write.
In the case of Fig. \ref{fig:schematic_operation}(b), the write and read MTJ is shared, i.e. each MTJ can be used to perform write or read operation depending on the bias voltage. The write/read MTJs are located at address "0x0" and "0x4", and thus multiple write and read can be achieved. Data can be written into these two addresses simultaneously with subsequent shift right operation. Consequently, the time required to write eight bits into a word is reduced by half compared to Fig \ref{fig:schematic_operation}(a). Furthermore, the write/read latency can also be improved by adding more write/read MTJs ports. By using the same write operation strategy, the time required to write a word can be further reduced by half in Fig. \ref{fig:schematic_operation}(c) compared to Fig. \ref{fig:schematic_operation}(b). Note more read ports, as in Fig. \ref{fig:schematic_operation}(b), \ref{fig:schematic_operation}(c), also improves read latency. 
The array organization of a skyrmion based cache memory is shown in Fig. \ref{fig:system}. The SWLs, R/WWLs are shared among all the multi-bit cells placed in a row, and BLs, SLs can be shared among all the multi-bit cells placed in a column. 
Considering the read/write MTJ is shared, the read and the write operations can be performed by turning on the access transistors (R/WWL) of the read/write MTJ, precharging the BL to $V_{WRITE}$ and $V_{READ}$, respectively. In addition, the shift operation can be performed by precharging the BL to $V_{SHIFT}$ and SL to GND. In this architecture, multiple words can be placed in the same row and accessed independently. Note that the decoder is used to select a multi-bit skyrmion cell in the array and the sense amplifier is used to detect the output signal to zero or one. \par  
Table \ref{tab:example} lists the bias voltage conditions (along with device dimensions) for write/shift/read/idle operations. 
A skyrmion can be nucleated (logic ONE) by injecting spin-polarized current through the write MTJ of diameter $20$ $nm$ with a current density of $2 \times 10^{13}$ $A/m^2$ for 0.5 $ns$. The write "1" (logic ONE) operation can be performed by turning ON the write access transistors, precharging BL to $V_{WRITE}$ and SL to GND. The idle operation is used to stabilize the magnetization of the nanotrack, and can be achieved by turning OFF all access transistors. We represent the absence of skyrmion to be logic ZERO and writing logic ZERO can be achieved just by shift operation.
To access the logic value of the stored data in distinct bits, shift operations are involved to move the target data to the position underneath the read MTJ. Shifting stored data in a multi-bit skyrmion cell can be accomplished by turning ON the access transistors of the shift ports and precharging BL and SL to appropriate operating voltage. In order to shift the stored data right, we precharge the BL to $V_{SHIFT}$ and SL to GND. 
On the other hand, to shift stored data in the opposite direction, the bias voltages of BL and SL are reversed.
The read operation can be performed by turning ON the access transistors of the read port, driving BL to $V_{READ}$ and SL to GND. The presence/absence of skyrmions can be detected by sensing the change of tunneling conductance from the read MTJ of diameter $20$ $nm$. The magnetoresistance ratio of the read MTJ is 200\%. However, since the average magnetization of a skyrmion is not anti-parallel to the fixed layer, a smaller magnetoresistance change can be obtained. This change is directly proportional to the diameter of the skyrmion, and is inversely proportional to the cross-sectional area of the MTJ. Thus, we chose the size of the read MTJ similar to the dimension of skyrmions in the nanotrack for better sensing. 
For accurate detection, the skyrmion should be located near the center of the read MTJ. However, the motion of a skyrmion might deviate from the center region of the nanotrack during the shift operation due to the Magnus force (which will be explained in the next section) \cite{thiele1973steady}. This deviation can be mitigated by relaxing skyrmions and allowing the skyrmions to move back to the center of the track through edge repulsion before executing the read operation (idle operation). Note that the required idle time to move skyrmions back to the center position is related to the deviation of the skyrmion from the center region, which depends on the drive current density.  
\begin{table}[!ht]
\centering
\begin{tabular}{|l|l|l|l|l|l|}
\hline
	  & RWL & WWL & SWL & BL & SL \\
\hline
Read  & $V_{DD}$ & 0 & 0 & $V_{READ}$  & 0 \\
\hline
Shift Left  & 0 & 0 & $V_{DD}$ & 0 & $V_{SHIFT}$ \\
\hline	
Shift Right & 0 & 0 & $V_{DD}$ & $V_{SHIFT}$ & 0 \\
\hline
Write & 0 & $V_{DD}$ & 0 & $V_{WRITE}$ & 0 \\
\hline
Idle  & 0 & 0 & 0 & 0 & 0 \\
\hline
\end{tabular}
\caption{\label{tab:example}Bias voltage conditions for various operations}
\end{table}
\subsection*{Motion of skyrmion}
\begin{figure}[!h]
\centering
\includegraphics[width=0.8\linewidth]{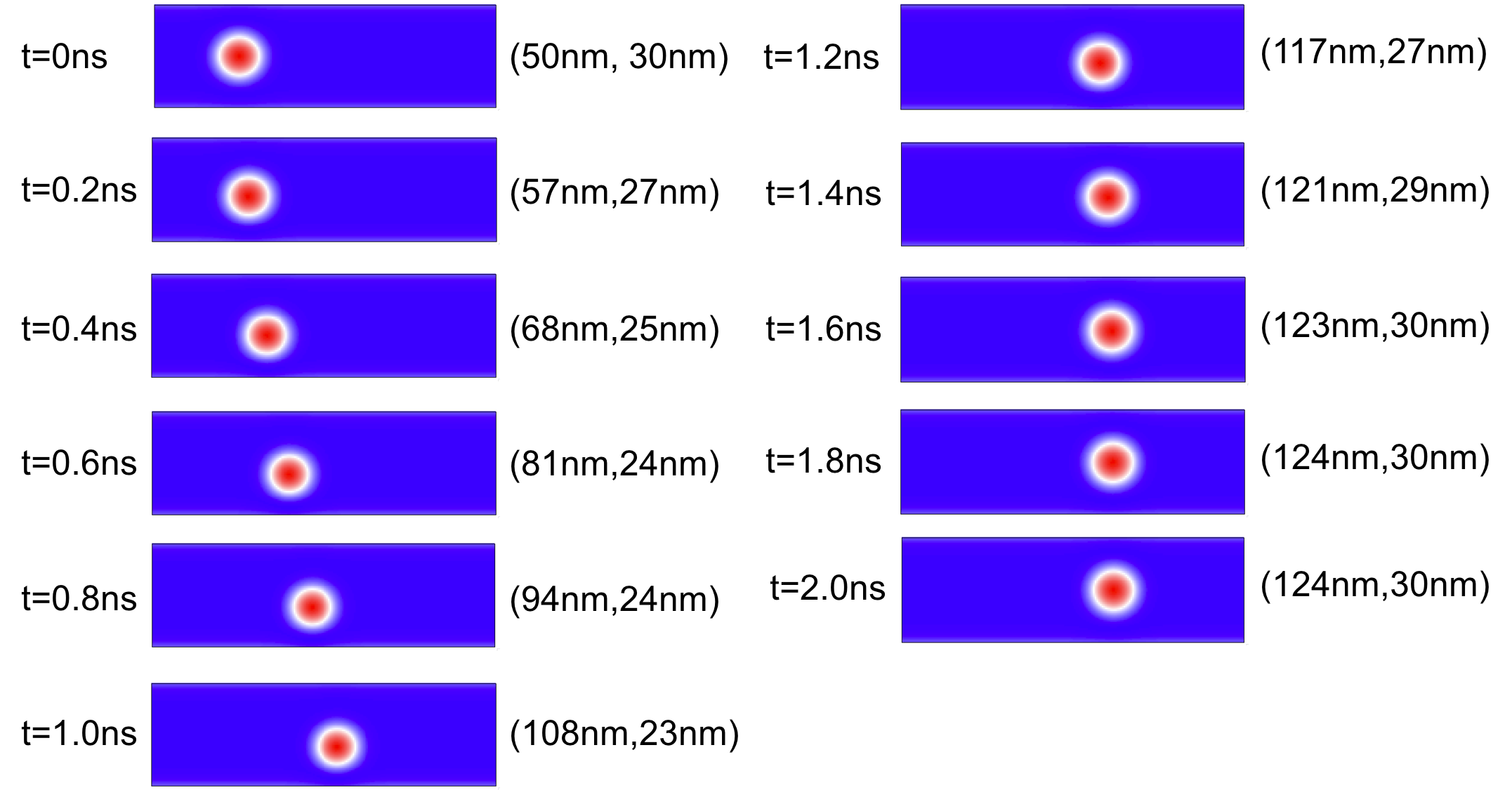}
\caption{The motion of skyrmion along the SHM layer current direction for a charge current density of $2.22\times 10^{11}$ $A/m^2$ through the SHM. The skyrmion moves toward the edge when an SHM current is applied for 1 $ns$, while it moves back to the center of nanotrack when the drive current is turned off.}
\label{fig:Operation}
\end{figure}

\begin{figure}[!ht]
\centering
\includegraphics[width=0.6\linewidth]{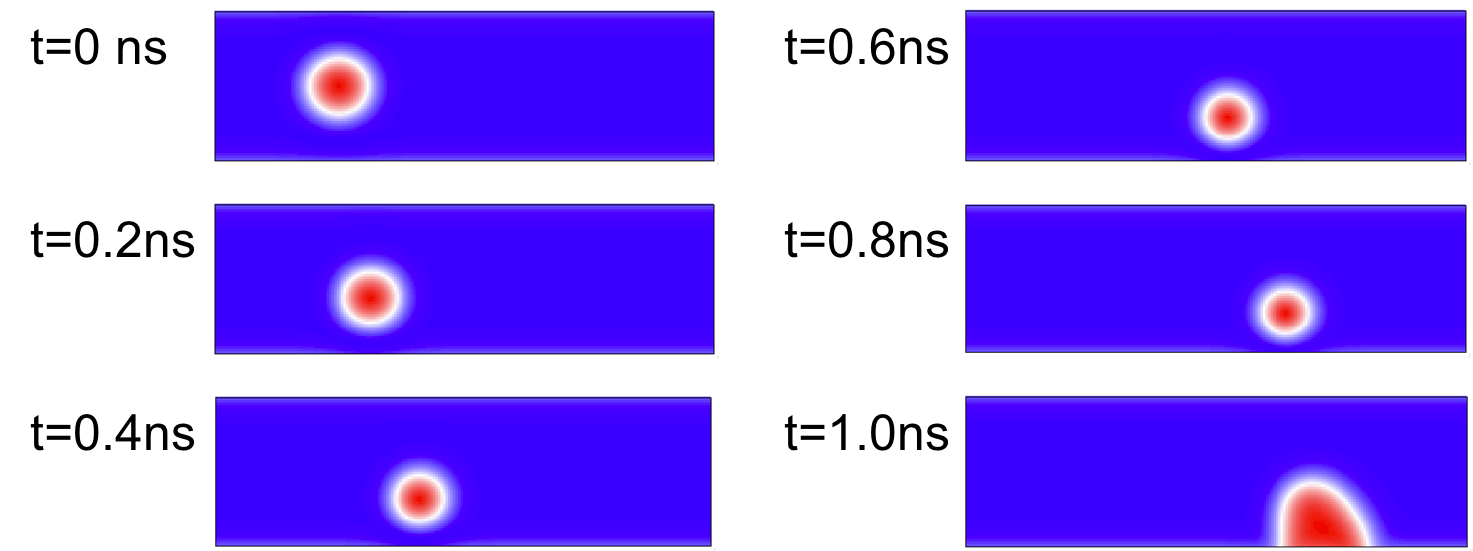}
\caption{Skyrmion annihilation from the edge of nanotrack caused by a current density of $4.44\times 10^{11}$ $A/m^2$ through the SHM}
\label{fig:Push_Out}
\end{figure}

In our proposed device structure, the skyrmion is driven by a vertical spin current generated from the charge current flowing through the SHM underlayer. The motion of the skyrmion can be well explained by the Theile's equation \cite{thiele1973steady}, 
\begin{equation}
G\times V_{d}-D \alpha v_{d} +j_{spin}=0 
\end{equation}
where $\alpha $ is the Gilbert damping constant, $G$ is gyromagnetic coupling, $D$ is the dissipative force tensor, $v_d$ is the drift velocity of a skyrmion, and $j_{spin}$ is the spin current induced by charge current flow through the SHM. The first term of eqn.(2) relates to the Magnus force caused by the interaction between the conduction electrons and the local magnetization. The longitudinal and transverse velocity can be written as
\begin{equation}
\begin{split}
v_d^x=\frac{\alpha D}{G^2+\alpha ^2 D^2} j_{spin}\\
v_d^y=\frac{G}{G^2+\alpha ^2 D^2} j_{spin}
\end{split}
\end{equation}
Figure \ref{fig:Operation} shows the trajectory of a skyrmion moving from the initial position to the next address position under a charge current density flowing through the SHM. Since a skyrmion undergoes a transverse velocity if $G\neq 0$, the motion of the skyrmion bends away from the drive current direction. Hence, shift and subsequent idle operations are required. The skyrmion is shifted right for 1 $ns$, followed by one cycle of idle operation (0.8 $ns$). In our simulations, ~0.8 ns idle time is required for the skyrmion to move back to the center along the width of the nanotrack. During the shift operation, the transverse motion of the skyrmion stops at a certain distance from the bottom edge due to the skyrmion-edge interaction. The distance from the bottom edge decreases as the SHM current density increases. Consequently, skyrmions are not annihilated from the edges unless a larger charge current density is applied along the SHM layer. 
In Fig. \ref{fig:Operation}, under a drive current density of $2.22\times 10^{11}$ $A/m^2$, the skyrmion stops at 23 $nm$ from the bottom edge. Fig. \ref{fig:Push_Out} shows the annihilation process of a skyrmion during shift operation in the nanotrack. Our simulation results show that the skyrmion annihilates in 1 $ns$ if the drive current density through the SHM is larger than $4.44\times 10^{11}$ $A/m^2$, which is ~2 times larger than the operation current density to shift a skyrmion to the next address position (Fig. \ref{fig:Operation}). As a result, for the shift operation, the SHM current density is being kept much below the level of annihilation.
\subsection*{Skyrmion packing density and Device Optimization}
\begin{figure}[!h]
\centering
\includegraphics[width=0.6\linewidth]{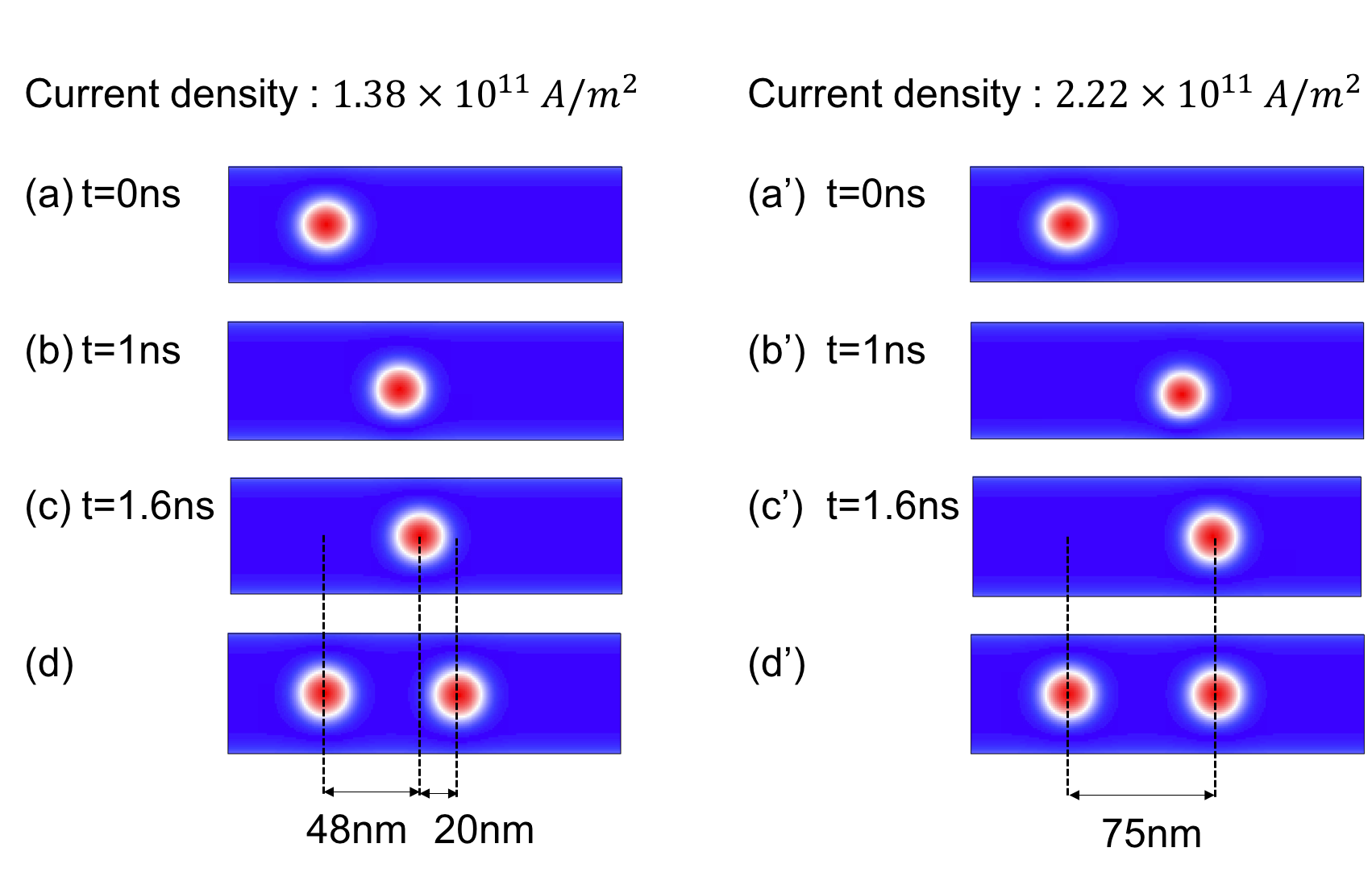}
\caption{(a)-(d) and (a')-(d') show a skyrmion is shifted rightward by 48 $nm$ and 75 $nm$ under an SHM current density of $1.38\times 10^{11}$ $A/m^2$ and $2.22\times 10^{11}$ $A/m^2$ for 1 $ns$, respectively, followed by 0.8 $ns$ relaxation time. (a) and (a') present a newly nucleated skyrmion after the write operation. (b) and (b') show the results after 1 $ns$ shift operation. (c) and (c') are the results of 0.8 $ns$ relaxation time after turning OFF shift current. (d) and (d') show the simulation result after a new skyrmion is nucleated underneath the write MTJ.}
\label{fig:Repulsive_Force}
\end{figure}

To ensure reliable read and write operation, the distance between consecutive skyrmion bits has to be consistent after each cycle of shift and idle operations. Furthermore, previously stored skyrmion bits should not be influenced by a newly nucleated skyrmion during the nucleation operation.
Consequently, reliable spacing between consecutive bits is required to alleviate the influence of repulsive forces from neighboring skyrmions. The material parameters used in our simulation were taken from ref. [3] and are shown in Table \ref{tab:example}.
In Fig. \ref{fig:Repulsive_Force}, a skyrmion is shifted rightward by 48 $nm$ and 75 $nm$ under an SHM current density of $1.38\times 10^{11}$ $A/m^2$ and $2.22\times 10^{11}$ $A/m^2$ for 1 $ns$, respectively, followed by 0.8 $ns$ relaxation time in each case.
When the spacing between skyrmions is 48 $nm$ (Fig. \ref{fig:Repulsive_Force}(a)-(d)), the stored skyrmion bit experiences repulsive force from the newly nucleated skyrmion, and thus the existing skyrmion moves 20 $nm$ right when a new skyrmion is nucleated. However, when the spacing between skyrmions is increased to 75 $nm$ (Fig. \ref{fig:Repulsive_Force}(a')-(d')), the stored skyrmion bit does not get affected, since the spacing is large enough to prevent any repulsive interaction from the newly nucleated skyrmion. Hence, a spacing of at least 75 nm between skyrmion bits is required (for the device dimensions and material parameters given in Table \ref{tab:example}) for reliable operation. It has been proposed that the size of skyrmions affects the reliable spacing between consecutive bits. 
The repulsive force between the two skyrmions can be described by \cite{lin2013particle}
\begin{equation}
F_{SS}=K_{1}(\frac{d\sqrt{H_{k}A}}{D})\times (\frac{A}{t})
\end{equation}
where $d$ is the distance between two center of skyrmions, $K_{1}$ is the modified Bessel function, $H_{K}$ is the perpendicular anisotropy, and A is the exchange stiffness, and $t$ is the thickness of the nanotrack.  \par
The size of a skyrmion affects the tunneling conductance of the read MTJ, the current required to move the skyrmion, and the packing density of the proposed cache design. The tunneling conductance of the read MTJ increases with increasing radius of skyrmion, and thus a larger read voltage swing can be obtained between the read and the reference MTJ. It has been discussed that the skyrmion velocity depends linearly on the size of the skyrmion \cite{manipulating}. Larger skyrmions couple more strongly with spin orbit torque, and thus faster motion can be induced. Therefore, a larger skyrmion can be shifted by a lower drive current density. However, the packing density reduces since the reliable spacing between two consecutive skyrmions increases for larger skyrmions \cite{zhang2015skyrmion}. The size of a skyrmion can be tuned by extrinsic parameter like external magnetic field, intrinsic parameters (such as DMI strength, PMA anisotropy constant), or by changing the width of the nanotrack. By applying an external magnetic field in the same direction as the polarization vector at the center of the skyrmion, a larger skyrmion can be obtained. In contrast, a skyrmion shrinks if the direction of the external magnetic field is opposite to the polarization vector at the skyrmion’s center. Similarly, the skyrmion size can be enlarged by increasing the DMI strength, or reducing the PMA anisotropy. Also, relaxation in the width of the nanotrack can result in larger skyrmion due to lower edge confinement effects.\par

\section*{Discussion}
\begin{figure}[!h]
\centering
\includegraphics[width=0.8\linewidth]{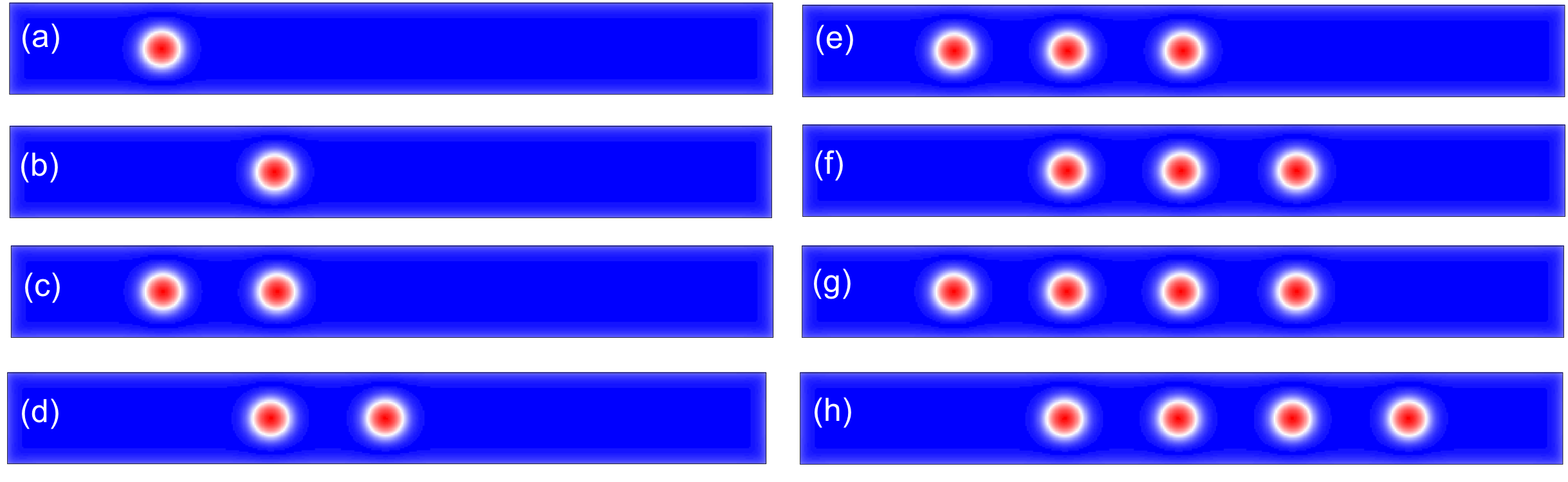}
\caption{Motion of skyrmions in our proposed device structure with reliable spacing between consecutive skyrmions}
\label{fig:4sk_Operation}
\end{figure}

We have proposed a skyrmion based cache memory in a 0.4 $nm$ thick nanotrack placed on top of a $700\times 60 \times 3$ $nm^3$ SHM. An MTJ of diameter 20 $nm$, similar to the size of the skyrmions in the nanotrack, has been used in the simulations. 
Fig. \ref{fig:4sk_Operation} shows the simulation results after several cycles of nucleation, shift, and idle operations for our proposed cache memory. 
By injecting a spin-polarized current density of $2 \times 10^{13}$ $A/m^2$ through the write MTJ for 0.5 $ns$ followed by 0.5 $ns$ idle time, a stable skyrmion can be nucleated (Fig. \ref{fig:4sk_Operation}(a)). Subsequently, an SHM current density of $2.22\times 10^{11}$ $A/m^2$ is used to shift the existing skyrmion rightward to the next position which is 75 $nm$ from the initial position (Fig. \ref{fig:4sk_Operation}(b)). 
Depending on the bias conditions, skyrmions in the nanotrack can be shifted in either direction. In order to write a sequence of data bits into the nanotrack, we shift skyrmions right after nucleation. 
Note that the drive current density should be lower than the current required to annihilate skyrmions from the edge ($4.44\times 10^{11}$ $A/m^2$ in our simulation). 
However, it should be higher than the minimum current density ($2.22\times 10^{11}$ $A/m^2$ in our simulation) to move each skyrmion at the same speed. 
In other words, the distance that the drive current density shifts the existing skyrmion from the initially nucleated position to the next position has to be large enough to mitigate the repulsive force from neighboring skyrmions. In our simulation, we find that at least 75 $nm$ is far enough to ensure that skyrmions do not experience repulsive force from the neighboring skyrmions. Hence, each skyrmion moves at a consistent speed under the injection of a vertical spin polarized current (as shown in Fig. \ref{fig:4sk_Operation}(d), \ref{fig:4sk_Operation}(f), \ref{fig:4sk_Operation}(h)). Also, as shown in Fig. \ref{fig:4sk_Operation}(c), \ref{fig:4sk_Operation}(e), \ref{fig:4sk_Operation}(g), the already existing skyrmions do not get affected by the newly nucleated skyrmion. The read operation can be performed after shifting and relaxing the target storage bit beneath the read MTJ. From our simulations we found that a voltage swing of 0.108 volt can be sensed between the read and the reference MTJ by pulling up BL voltage to 0.8 volt and SL to GND. \par 
Compared with DW based cache memory, the current density required to move skyrmions is much lower than a domain wall. Also, the flexibility of a skyrmion allows it to be less pinned by defects than a domain wall, making skyrmion based memory more reliable. Since the average magnetization of a skyrmion is not anti-parallel to the fixed layer, a smaller magnetoresistance change would be detected compared to the DW based one. Thus, the size of the read MTJ similar to the dimension of skyrmions is chosen to obtain a higher magnetoresistance change. 

\section*{Methods}
\textbf{Micromagnetic modeling.} The micromagnetic simulations are performed using graphics-processing-unit-based tool Mumax3 \cite{najafi2009proposal,vansteenkiste2014design}. The magnetization dynamics of magnetic skyrmions driven by vertical current can be expressed by 
\begin{equation}
\begin{split}
\tau & = \frac{\gamma}{1+\alpha^2}(m\times H_{eff}+\alpha(m\times(m \times H_{eff})))+\tau_{SL} \\
\tau_{SL} &=\beta\frac{\epsilon-\alpha \epsilon\prime}{1+\alpha^2}(m \times (m_p \times m))-\beta \frac{\epsilon\prime-\alpha \epsilon}{1+\alpha^2} m \times m_p \\
\beta & =\frac{j_z\hbar}{M_{sat}ed}  \\
\epsilon & =\frac{P\Lambda^2}{(\Lambda^2+1)+(\Lambda^2-1)(m\cdot m_p)} \\
\end{split}
\end{equation}
where $m$ is the normalized magnetization vector, $m_p$ is the fixed layer polarization, $\gamma$ is Gilbert gyromagnetic ratio , $\alpha$ is Gilbert damping parameter, $H_{eff}$ is the effective field, $j_z$ is the current density along the z axis, $M_{sat}$ is the saturation magnetization, e is the elementary charge, d is the skyrmion layer thickness, P is the polarization of conduction electron, the Slonczewski $\Lambda$ parameter characterizes the spacer layer, and $\epsilon \prime$ is the secondary spin transfer term. 
The material parameters used in our simulations correspond to Co/Pt multilayers \cite{Co/Pt_barman2007ultrafast,Co/Pt_metaxas2007creep}, and are shown in Table \ref{tab:example2}. We consider a 0.4 $nm$ thick Co nanotrack with perpendicular magnetic anisotropy on a 3 $nm$ Pt substrate inducing DMI. The sample is discretized into an element size of $1 \times 1\times 0.4$ $nm^3$. \par
The Non-equilibrium Green's Function (NEGF) based spin transport simulation has been used in order to obtain the resistance of the MTJ. The charge current ($I_e$) flowing through the SHM and the corresponding spin current ($I_s$) are calculated using\cite{liu2011spin} 
\begin{equation}
I_s=\theta_{sh} \frac{A_{MTJ}}{A_{SHM}}I_e
\end{equation}
where $A_{MTJ}$ and $A_{SHM}$ are the cross sectional areas of the MTJ and SHM, respectively, and $\theta_{sh}$ is the spin-Hall angle. The spin current from eqn.(6) is used to analyze the magnetic dynamics with the generalized LLGS equation. Magnetization dynamics simulations are performed using the Mumax3 platform.   

\begin{table}[!h]
\centering
\begin{tabular}{|l|l|}
\hline
Parameter & Value \\
\hline
Saturation magnetization ($M_{sat}$)  & $5.8\times 10^5 A/m$  \\
\hline
PMA anisotropy constant ($K_u$)  & $0.8\times 10^6J/m^3$ \\
\hline	
Exchange constant (A) & $1.5\times 10^{11}J/m $ \\
\hline
Dzyaloshinskii-Moriya interaction (DMI) strength ($D$) & $3.5 mJ/m^2$  \\
\hline
Gilbert damping constant ( $\alpha$ ) & 0.3  \\
\hline
Gilbert gyromagnetic ratio ( $\gamma$ ) & $2.211\times 10^5 m/As$  \\
\hline
Spin polarization ( $P$ ) & 0.4  \\
\hline
Nanotrack dimensions & $700 \times 60 \times 0.4$ $nm^3$ \\
\hline
SHM dimensions & $700 \times 60 \times 3$ $nm^3$ \\
\hline
MTJ diameter & $20nm$  \\
\hline
Spin Hall Angle ($\theta_{sh}$) & 0.07 \\
\hline
\end{tabular}
\caption{\label{tab:example2}Material parameters used for simulation}
\end{table}
\bibliography{reference}

\begin{thebibliography}{10}
\expandafter\ifx\csname url\endcsname\relax
  \def\url#1{\texttt{#1}}\fi
\expandafter\ifx\csname urlprefix\endcsname\relax\def\urlprefix{URL }\fi
\providecommand{\bibinfo}[2]{#2}
\providecommand{\eprint}[2][]{\url{#2}}

\bibitem{IBM_racetrack}
\bibinfo{author}{Parkin, S.~S.}, \bibinfo{author}{Hayashi, M.} \&
  \bibinfo{author}{Thomas, L.}
\newblock \bibinfo{title}{Magnetic domain-wall racetrack memory}.
\newblock \emph{\bibinfo{journal}{Science}} \textbf{\bibinfo{volume}{320}},
  \bibinfo{pages}{190--194} (\bibinfo{year}{2008}).

\bibitem{tapecache}
\bibinfo{author}{Venkatesan, R.} \emph{et~al.}
\newblock \bibinfo{title}{Tapecache: a high density, energy efficient cache
  based on domain wall memory}.
\newblock In \emph{\bibinfo{booktitle}{Proceedings of the 2012 ACM/IEEE
  international symposium on Low power electronics and design}},
  \bibinfo{pages}{185--190} (\bibinfo{organization}{ACM},
  \bibinfo{year}{2012}).

\bibitem{nucleation}
\bibinfo{author}{Sampaio, J.}, \bibinfo{author}{Cros, V.},
  \bibinfo{author}{Rohart, S.}, \bibinfo{author}{Thiaville, A.} \&
  \bibinfo{author}{Fert, A.}
\newblock \bibinfo{title}{Nucleation, stability and current-induced motion of
  isolated magnetic skyrmions in nanostructures}.
\newblock \emph{\bibinfo{journal}{Nature nanotechnology}}
  \textbf{\bibinfo{volume}{8}}, \bibinfo{pages}{839--844}
  (\bibinfo{year}{2013}).

\bibitem{defect_in_DWM}
\bibinfo{author}{Thiaville, A.}, \bibinfo{author}{Nakatani, Y.},
  \bibinfo{author}{Miltat, J.} \& \bibinfo{author}{Suzuki, Y.}
\newblock \bibinfo{title}{Micromagnetic understanding of current-driven domain
  wall motion in patterned nanowires}.
\newblock \emph{\bibinfo{journal}{EPL (Europhysics Letters)}}
  \textbf{\bibinfo{volume}{69}}, \bibinfo{pages}{990} (\bibinfo{year}{2005}).

\bibitem{dzyaloshinskyDMI}
\bibinfo{author}{Dzyaloshinsky, I.}
\newblock \bibinfo{title}{A thermodynamic theory of “weak” ferromagnetism
  of antiferromagnetics}.
\newblock \emph{\bibinfo{journal}{Journal of Physics and Chemistry of Solids}}
  \textbf{\bibinfo{volume}{4}}, \bibinfo{pages}{241--255}
  (\bibinfo{year}{1958}).

\bibitem{moriyaDMI}
\bibinfo{author}{Moriya, T.}
\newblock \bibinfo{title}{New mechanism of anisotropic superexchange
  interaction}.
\newblock \emph{\bibinfo{journal}{Physical Review Letters}}
  \textbf{\bibinfo{volume}{4}}, \bibinfo{pages}{228} (\bibinfo{year}{1960}).

\bibitem{fert2013skyrmions}
\bibinfo{author}{Fert, A.}, \bibinfo{author}{Cros, V.} \&
  \bibinfo{author}{Sampaio, J.}
\newblock \bibinfo{title}{Skyrmions on the track}.
\newblock \emph{\bibinfo{journal}{Nature nanotechnology}}
  \textbf{\bibinfo{volume}{8}}, \bibinfo{pages}{152--156}
  (\bibinfo{year}{2013}).

\bibitem{nagaosa2013topological}
\bibinfo{author}{Nagaosa, N.} \& \bibinfo{author}{Tokura, Y.}
\newblock \bibinfo{title}{Topological properties and dynamics of magnetic
  skyrmions}.
\newblock \emph{\bibinfo{journal}{Nature nanotechnology}}
  \textbf{\bibinfo{volume}{8}}, \bibinfo{pages}{899--911}
  (\bibinfo{year}{2013}).

\bibitem{muhlbauer2009skyrmion}
\bibinfo{author}{M{\"u}hlbauer, S.} \emph{et~al.}
\newblock \bibinfo{title}{Skyrmion lattice in a chiral magnet}.
\newblock \emph{\bibinfo{journal}{Science}} \textbf{\bibinfo{volume}{323}},
  \bibinfo{pages}{915--919} (\bibinfo{year}{2009}).

\bibitem{yu2010real}
\bibinfo{author}{Yu, X.} \emph{et~al.}
\newblock \bibinfo{title}{Real-space observation of a two-dimensional skyrmion
  crystal}.
\newblock \emph{\bibinfo{journal}{Nature}} \textbf{\bibinfo{volume}{465}},
  \bibinfo{pages}{901--904} (\bibinfo{year}{2010}).

\bibitem{yu2011near}
\bibinfo{author}{Yu, X.} \emph{et~al.}
\newblock \bibinfo{title}{Near room-temperature formation of a skyrmion crystal
  in thin-films of the helimagnet fege}.
\newblock \emph{\bibinfo{journal}{Nature materials}}
  \textbf{\bibinfo{volume}{10}}, \bibinfo{pages}{106--109}
  (\bibinfo{year}{2011}).

\bibitem{heinze2011spontaneous}
\bibinfo{author}{Heinze, S.} \emph{et~al.}
\newblock \bibinfo{title}{Spontaneous atomic-scale magnetic skyrmion lattice in
  two dimensions}.
\newblock \emph{\bibinfo{journal}{Nature Physics}}
  \textbf{\bibinfo{volume}{7}}, \bibinfo{pages}{713--718}
  (\bibinfo{year}{2011}).

\bibitem{liu2012spin}
\bibinfo{author}{Liu, L.} \emph{et~al.}
\newblock \bibinfo{title}{Spin-torque switching with the giant spin hall effect
  of tantalum}.
\newblock \emph{\bibinfo{journal}{Science}} \textbf{\bibinfo{volume}{336}},
  \bibinfo{pages}{555--558} (\bibinfo{year}{2012}).

\bibitem{finocchio2013switching}
\bibinfo{author}{Finocchio, G.}, \bibinfo{author}{Carpentieri, M.},
  \bibinfo{author}{Martinez, E.} \& \bibinfo{author}{Azzerboni, B.}
\newblock \bibinfo{title}{Switching of a single ferromagnetic layer driven by
  spin hall effect}.
\newblock \emph{\bibinfo{journal}{Applied Physics Letters}}
  \textbf{\bibinfo{volume}{102}}, \bibinfo{pages}{212410}
  (\bibinfo{year}{2013}).

\bibitem{thiele1973steady}
\bibinfo{author}{Thiele, A.}
\newblock \bibinfo{title}{Steady-state motion of magnetic domains}.
\newblock \emph{\bibinfo{journal}{Physical Review Letters}}
  \textbf{\bibinfo{volume}{30}}, \bibinfo{pages}{230} (\bibinfo{year}{1973}).

\bibitem{manipulating}
\bibinfo{author}{Ding, J.}, \bibinfo{author}{Yang, X.} \& \bibinfo{author}{Zhu,
  T.}
\newblock \bibinfo{title}{Manipulating current induced motion of magnetic
  skyrmions in the magnetic nanotrack}.
\newblock \emph{\bibinfo{journal}{Journal of Physics D: Applied Physics}}
  \textbf{\bibinfo{volume}{48}}, \bibinfo{pages}{115004}
  (\bibinfo{year}{2015}).

\bibitem{zhang2015skyrmion}
\bibinfo{author}{Zhang, X.} \emph{et~al.}
\newblock \bibinfo{title}{Skyrmion-skyrmion and skyrmion-edge repulsions in
  skyrmion-based racetrack memory}.
\newblock \emph{\bibinfo{journal}{Scientific reports}}
  \textbf{\bibinfo{volume}{5}} (\bibinfo{year}{2015}).

\bibitem{liu2011spin}
\bibinfo{author}{Liu, L.}, \bibinfo{author}{Moriyama, T.},
  \bibinfo{author}{Ralph, D.} \& \bibinfo{author}{Buhrman, R.}
\newblock \bibinfo{title}{Spin-torque ferromagnetic resonance induced by the
  spin hall effect}.
\newblock \emph{\bibinfo{journal}{Physical review letters}}
  \textbf{\bibinfo{volume}{106}}, \bibinfo{pages}{036601}
  (\bibinfo{year}{2011}).

\end{thebibliography}
\section*{Acknowledgments}
The work was supported in part by, Center for Spintronic Materials, Interfaces, and Novel Architectures (C-SPIN), a MARCO and DARPA sponsored StarNet center, by the Semiconductor Research Corporation, the National Science Foundation, Intel Corporation and by the National Security Science and Engineering Faculty Fellowship. 
\section*{Author contributions statement}
M.C. and K.R coordinated the project. M.C. performed the micromagnetic simulations supervised by K.R. All authors interpreted the results and prepared the manuscript.
\section*{Competing financial interests}
The authors declare no competing financial interests.
\end{document}